\documentclass[letter]{aa}  

\usepackage{graphicx}
\usepackage{txfonts}
\usepackage{color}
\begin{document}

   \title{
  Long-term millimeter VLBI monitoring of M\,87 with KVN \\
  at milliarcsecond resolution: nuclear spectrum
  }

   \subtitle{}

   \author{
        Jae-Young Kim\inst{1} 
   \and Sang-Sung Lee\inst{2,3}
   \and Jeffrey A. Hodgson\inst{2}
   \and Juan-Carlos Algaba\inst{2,4}
   \and Guang-Yao Zhao\inst{2}
   \and Motoki Kino\inst{2}
   \and Do-Young Byun\inst{2}
   \and Sincheol Kang\inst{2,3}
          }

   \institute{
   Max-Planck-Institut f\"ur Radioastronomie, Auf dem H\"ugel 69, D-53121 Bonn, Germany  \\ 
   \email{jykim@mpifr-bonn.mpg.de} 
   \and
   Korea Astronomy and Space Science Institute, 776 Daedeokdae-ro, Yuseong-gu, Daejeon, 30455, Korea
   \and
   Korea University of Science and Technology, 217 Gajeong-ro, Yuseong-gu, Daejeon 34113, Korea
   \and
   Department of Physics and Astronomy, Seoul National University, 1 Gwanak-ro, Gwanak-gu, Seoul 08826, Korea
             }

   \date{Received ; accepted }

 
  \abstract{
We study the centimeter- to millimeter-wavelength synchrotron spectrum
of the core of the radio galaxy M\,87 at $\lesssim0.8\,{\rm mas}~\sim110R_{s}$ spatial scales
using four years of fully simultaneous, multi-frequency VLBI data obtained by the Korean VLBI Network (KVN).
We find a core spectral index $\alpha$ of $\gtrsim-0.37$ ($S\propto \nu^{+\alpha}$) between 22\,GHz and 129\,GHz.
By combining resolution-matched flux measurements from the Very Long Baseline Array (VLBA) at 15\,GHz
and taking the Event Horizon Telescope (EHT) 230\,GHz core flux measurements in epochs 2009 and 2012 as lower limits,
we find evidence of a nearly flat core spectrum across 15\,GHz and 129\,GHz,
which could naturally connect the 230~GHz VLBI core flux.
The extremely flat spectrum is a strong indication that the jet base does not consist of a simple homogeneous plasma,
but of inhomogeneous multi-energy components, with at least one component with the turn-over frequency $\gtrsim100$\,GHz.
The spectral shape can be qualitatively explained if both the strongly (compact, optically thick at $>$100\,GHz) and 
the relatively weakly magnetized (more extended, optically thin at $<$100\,GHz) plasma components are 
colocated in the footprint of the relativistic jet.}


   \titlerunning{KVN M\,87 VLBI monitoring}
   
   \authorrunning{J.-Y. Kim et al.}
               
   \maketitle
%

\section{Introduction}\label{sec:intro}

Located at 16.7\,Mpc \citep{bird10} from the Milky Way,
M\,87 is one of the nearest radio galaxies that contains an extraordinary
massive central black hole $M_{\rm BH}\approx6.1\times10^{9}M_{\odot}$ \citep{gebhardt11}
\footnote{
Some studies suggest a twice lower value of $M_{\rm BH}$ (e.g., \citealt{walsh13}).
In order to be consistent with other VLBI studies of M\,87, we adopt $M_{\rm BH}=6.1\times10^{9}M_{\odot}$.}.
The close distance and the large $M_{\rm BH}$ provides an excellent spatial resolution
(1 milli-arcsecond$\sim0.08$\,pc or $\sim$140 Schwarzschild radii ($R_{\rm s}$)).
This makes M\,87 a more than suitable target for observing and studying 
physical conditions in the relativistic jet launching, collimation, and acceleration region
by very long baseline interferometry (VLBI) observations 
\citep{kovalev07,ly07,hada11,asada12,doeleman12,hada16,mertens16,kim16,walker16}.

The electron energy distribution and the magnetic field strength near the central engine
are critical elements in relativistic jet launching models (e.g., \citealt{bz,bp}).
In particular, a stronger magnetic field helps effective jet formation (see \citealt{yuan14} for a review).
These models prefer a significantly inverted or at least flat synchrotron spectrum up to 
$>100$\,GHz \citep{broderick09,kino15,punsly17}.
A high turn-over frequency like this would imply a strong magnetic field strength $B\sim100$~G in the jet base \citep{kino15}.
However, the previous studies of M\,87 are based on data from vastly different observing times and spatial scales.

For this reason, simultaneous multi-frequency VLBI observations of M\,87 in the millimeter regime 
can provide better observational constraints for the theoretical models. 
However, such an observation has been challenging.
One of the main difficulties has been the limited frequency coverage offered by most VLBI observatories ($\leq86$~GHz=3.5~mm; \citealt{hada16,kim16}).
Recent Event Horizon Telescope (EHT) observations studied M\,87 at 230\,GHz (1.3~mm; e.g., \citealt{doeleman12}),
but the large frequency gap between 86 and 230~GHz and 
technical difficulties for submillimeter VLBI remain challenging.
Furthermore, the VLBI core of M\,87 varies in flux
on timescales of at least weeks, especially during the ejection of a new VLBI feature from the core region (e.g., \citealt{acciari09,hada14}).
The long-term variability properties in this frequency range are also poorly known.

In this letter, we present a study of the M\,87 VLBI core spectrum
based on the fully simultaneously measured core flux at 22, 43, 86~GHz and up to 129~GHz by the Korean VLBI Network (KVN) over the past four years.
We especially investigate the core spectrum at short millimeter-wavelengths and attempt to clarify
whether the nuclear region has a significantly inverted or steep spectrum.


\section{Observations and data processing} \label{sec:method}


\begin{table*}[!t]
\caption{Observing epochs and the M\,87 core model-fit parameters.}
\label{tab:1}
\centering
\begin{tabular}{c c c c c c c c c c}
\hline\hline
Epoch &
MJD &
$\nu_{\rm obs}$ &
$S_{\nu}$ &
$S_{\rm peak}$ &
$d$ &
$\sigma_{\rm rms}$ &
$B_{\rm maj}$ &
$B_{\rm min}$ &
$B_{\rm PA}$ \\
(yyyy-mm-dd) &
 &
(GHz) &
(Jy) &
(Jy/beam) &
(mas) &
(mJy/beam) &
(mas) &
(mas) &
(deg) \\
(1) &
(2) &
(3) &
(4) &
(5) &
(6) &
(7) &
(8) &
(9) &
(10) \\
\hline
2012-12-04 & 56265 & 21.7 & 1.7$\pm$0.2 & 1.7$\pm$0.1 & $<0.31$ & 10.0 & 5.70 & 3.22 & $-$68.0 \\
 &  & 43.4 & 1.3$\pm$0.2 & 1.3$\pm$0.1 & $<0.17$ & 9.0 & 2.84 & 1.61 & $-$66.3 \\
 &  & 86.8 & 1.1$\pm$0.2 & 1.1$\pm$0.1 & 0.18$\pm$0.02 & 13.9 & 1.39 & 0.83 & $-$67.0 \\
2013-01-16 & 56308 & 21.7 & 1.9$\pm$0.2 & 1.8$\pm$0.1 & 0.91$\pm$0.06 & 8.0 & 5.33 & 3.32 & $-$79.5 \\
\hline
\end{tabular}
\tablefoot{
The columns show
(1) the mean observing epoch,
(2) the corresponding MJD,
(3) the central observing frequency,
(4) the core flux density,
(5) the peak intensity,
(6) the core FWHM size,
(7) the image rms noise level after subtracting the Gaussian model,
(8) and (9) the major and minor axis of the elliptical observing beam,
and
(10) the beam position angle in degree (zero to North, increase in CCW).
Only the first four lines of the whole table are shown here for illustration,
and the entire table in electronic form is available in the online version.
}
\end{table*}

M\,87 was monitored regularly by the KVN between Dec 2012 and Dec 2016 (31 epochs)
as one of the samples included in the interferometric monitoring of gamma-ray bright AGN (iMOGABA) program 
(e.g., \citealt{lee16})\footnote{
http://radio.kasi.re.kr/sslee/
}.
The observations were performed simultaneously at four observing frequencies $\nu$ of
21.700 - 21.764\,GHz (K band), 
43.400 - 43.464\,GHz (Q band), 
86.800 - 86.864\,GHz (W band), and 
129.300 - 129.364\,GHz (D band). 
Each band has 64~MHz of total bandwidth.
The source was observed in left-circular polarization.
Details on the iMOGABA program (data acquisition, correlation, post-processing,
and calibration, especially concerning the frequency-phase transfer) are provided elsewhere
(see \citealt{lee15,algaba15,hodgson16,lee16}).
Owing to the frequency-phase transfer technique, we detected fringes at the higher frequency bands (e.g., 86GHz and 129GHz). 
After detection of the fringes, we averaged the calibrated data in 30\,s intervals at K and Q band, and 
in 10\,s intervals at W and D band to avoid amplitude loss from decoherence.
Occasional station or receiver problems and poor weather conditions 
led to unreliable amplitude and large residual phase errors in a certain portion of the data, especially at D band.
We excluded these data in the further analysis.

We used the \textsc{Difmap} package \citep{difmap} for imaging using \textsc{clean} and
phase self-calibration loops, applying natural weighting (no amplitude self-calibration because there are only three KVN stations).
An asymmetric and extended jet structure was revealed at 22 and 43~GHz by excess of fluxes in the dirty map and 
also by non-zero closure phases.
However, the 86 and 129~GHz data did not show a clear signature of extended jet emission
(see Fig. 2 in \citealt{lee16} for typical visibility distributions and source images).
We fitted a circular Gaussian to the self-calibrated visibilities using 
the \textsc{Modelfit} procedure in \textsc{Difmap} to estimate the total flux density $S_{\rm tot}$ (in Jy),
the FWHM size $d$ (in mas),
and the peak flux density $S_{\rm peak}$ (in mJy/beam) of the core.
We estimated uncertainties in the model-fit parameters and practical resolution limit of the array
by following \cite{lee16}, accounting for the effect of the finite signal-to-noise ratio (S/N)
that varies from session to session.
At D band, the systematic flux uncertainty can be as large as 30\% \citep{lee16} because of large residual phase errors.
Therefore, we take the 30\% as a conservative flux uncertainty at 129~GHz.
We also investigated possible amplitude loss at D band due to the decoherence
by averaging the data over different timescales (from 2\,s to 60\,s)
and measuring the peak intensity in the clean image.
We found that the 10\,s averaging can cause $\lesssim10$\% of amplitude loss.
This is not significant compared to the 30\% flux uncertainty.
Thus, we ignore the decoherence in the following discussions.


\section{Results and analysis}\label{sec:results}

\subsection{Core model-fit properties}\label{subsec:result}

In Table \ref{tab:1} we show a summary of the observations and the results of the core flux measurements.
We find that 45\%, 65\%, 76\%, and 38\% of all the model components are 
spatially resolved at K, Q, W, and D band, respectively.
The mean FWHM sizes of the resolved components are
$0.8\pm0.3$, $0.55\pm0.11$, $0.42\pm0.13$, and $0.23\pm0.09$\,mas 
at K, Q, W, and D bands, where the uncertainties correspond to the standard deviation of each distributions.
Higher resolution VLBI observations show that the source consists of
a more compact core and an extended, complex jet \citep{kovalev07,ly07,hada16,kim16}.
Therefore, the KVN multi-frequency observations probe synchrotron emission from the mixed structure
on scales smaller than $\sim0.8$~mas (equivalently $\sim110R_{s}$ projected linear size).

\begin{figure}
\centering
\includegraphics[width=0.45\textwidth]{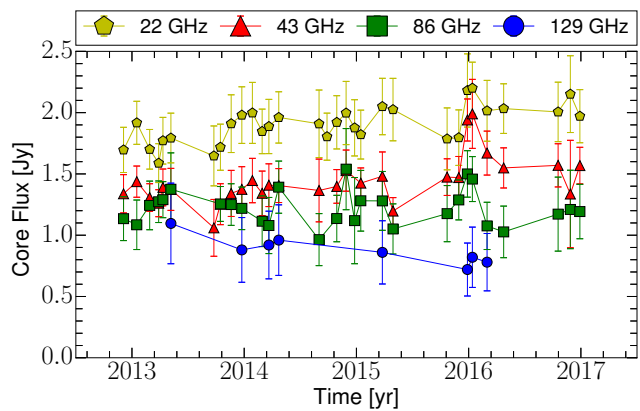}
\caption{
Four-frequency core flux light curves obtained by the KVN observations.
}
\label{fig:lc_flux}
\end{figure}

Our multi-frequency light curves are shown in Fig. \ref{fig:lc_flux}.
The core does not show significant flux variability over 1$\sigma$ during most of the observing epochs.
However, we note that the core flux increased from $\sim1.5$~Jy to $\sim2.0$~Jy at 43~GHz
and similarly at 22 and 86~GHz in the beginning of 2016. A study of this flux enhancement will be presented in a forthcoming publication.
The time-averaged core flux (including both resolved and unresolved components) are
$1.90\pm0.15$, $1.44\pm0.19$, $1.22\pm0.14$, and $0.88\pm0.11$\,Jy
at K, Q, W, and D band, respectively.
These values are broadly consistent with previously reported flux values of the M\,87 nuclear region on a similar spatial scale.
For instance, the KVN Q-band flux ($\sim1.4$\,Jy) is comparable to
the M\,87 core flux measured within 1.2~mas from the intensity peak by the Very Long Baseline Array (VLBA) 43~GHz observations ($\sim1.2$~Jy; \citealt{acciari09}).
%


\subsection{Synchrotron spectrum analysis}\label{subsec:analysis}

\begin{figure}
\centering
\includegraphics[width=0.45\textwidth]{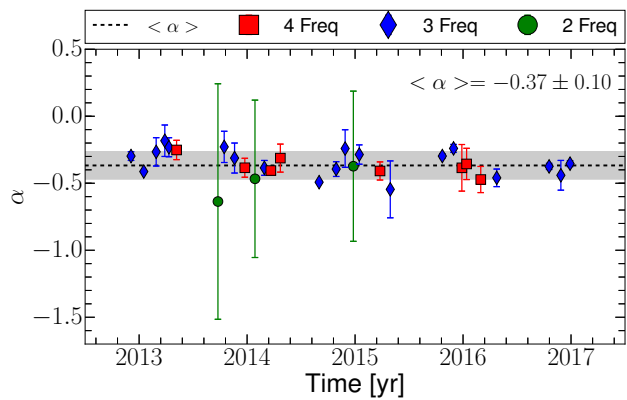}
\caption{
Spectral index light curve ($S\propto\nu^{+\alpha}$).
Different symbols in each epoch denote the different number of frequencies available for the power-law fit. 
The broken line and the shaded region denote the mean and standard deviation of $\alpha$, respectively.
}
\label{fig:lc_spix}
\end{figure}

\begin{figure*}
\centering
\includegraphics[width=0.81\textwidth]{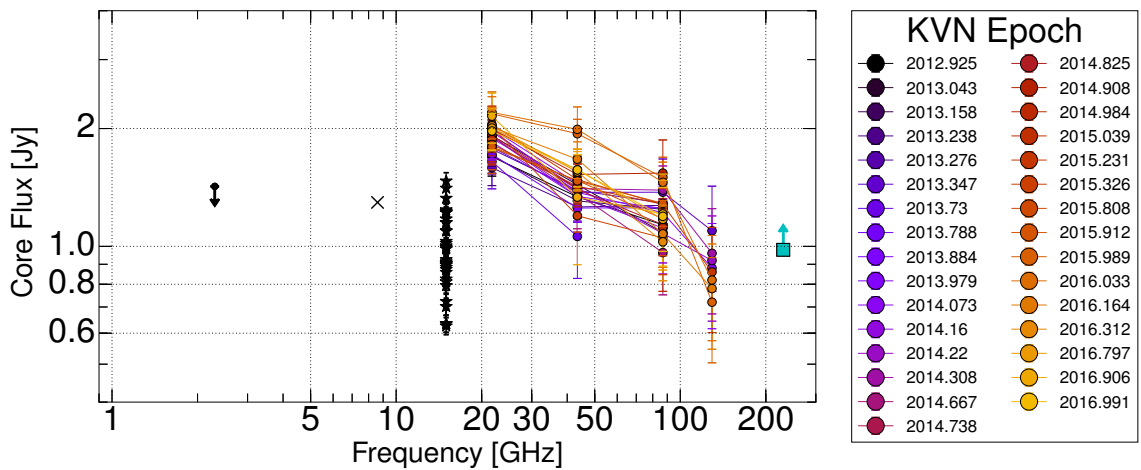}
\caption{
Radio spectrum of the M\,87 core region in log-log scale.
Different KVN observing epochs are indicated by different colors in the legend.
The arrows indicate upper and lower limits.
References of other multi-frequency flux information (in order of increasing radio frequency):
\cite{pushkarev12,lister16,doeleman12,akiyama15}.
}
\label{fig:sed}
\end{figure*}

Based on the absence of a significant synchrotron peak between 22-129~GHz and given a total of four frequencies available,
we modeled the core spectrum by a single power-law,
that is, $S_{\rm tot} \propto \nu^{+\alpha}$ , where $\alpha$ is the optically thin spectral index (negative for optically thin plasma).
In each epoch, we used the core flux from as many available frequencies as possible.
In Fig. \ref{fig:lc_spix} we show the spectral index light curve
and in Fig. \ref{fig:sed} the core flux versus the frequency.
The spectral index shows some oscillatory pattern with a $\sim$1\,yr timescale 
and a weak trend of spectral steepening over a longer period.
By averaging the spectral index light curve, we obtained the mean spectral index $\langle\alpha\rangle=-0.37\pm0.10$, 
which corresponds to a mildly steep spectrum.

However, we note that the limited $(u,v)$ coverage of KVN could have a significant effect on the core spectral index measurement
because of the different array sensitivity to extended emission at different frequencies.
In order to study the effect of the $(u,v)$ coverage, we have performed a dedicated imaging simulation (see the appendix).
We find that an artificial spectral steepening of $\Delta\alpha\sim-0.2$ could be possible for the M\,87-like jet structure
only due to the $(u,v)$-coverage effect (equivalent to $\sim36$\% of flux decrease from 22 to 129~GHz).
Furthermore, M\,87 has a substantially extended jet, whose spectrum is quite steep 
already at $\gtrsim0.2$~mas core separation ($\alpha\lesssim-0.7$; \citealt{hovatta14,hada16}).
Therefore, the structure blending effect could make the KVN M\,87 core spectrum steeper
even if the intrinsic spectrum is nearly flat.
Considering the imaging simulation results,
we conclude that the true spectrum of the jet base region is substantially flat up to 129~GHz.

We verified this conclusion by comparing 
non-simultaneous M\,87 core flux information that is available from earlier publications.
Here we specifically paid attention to the different resolutions of different observations in order to
compare the fluxes on a comparable angular resolution.
First, \cite{pushkarev12} reported an 8.6~GHz core flux of $\sim1.3$~Jy and an FWHM size of $\sim0.6$~mas.
The authors also reported a 2.3~GHz core flux of $\sim1.4$~Jy and an FWHM size of $\sim2.6$~mas along the major axis 
of an elliptical Gaussian.
Second, we made use of the M\,87 core information provided by the MOJAVE program 
(\citealt{lister16}; 34 epochs of VLBA 15\,GHz).
The time-averaged core flux and the FWHM size are $0.99\pm0.22$~Jy and 0.26$\pm$0.04\,mas, respectively.
Finally, we refer to the 230\,GHz EHT observations in epochs 2009 \citep{doeleman12} and 2012 \citep{akiyama15}.
At this frequency, an ultra-compact plasma ($\sim40~\mu$as)
has a flux of $\sim1$~Jy.
We take the 2.3~GHz flux as the upper limit since the 2.3~GHz size is an order-of-magnitude larger with regard to the KVN results.
For similar reasons, we regard the EHT flux as the lower limit at 230~GHz.

We find that the global radio spectrum of the core is considerably flat 
without a clear signature of inverted or steep spectra (Fig. \ref{fig:sed}).
In particular, the long-term 15~GHz core flux is quite comparable to our four-year measurements made by the KVN at 86 and 129\,GHz.
Since the angular resolution of the KVN at short millimeter wavelengths is comparable to 
that of VLBA at 15\,GHz,
we can calculate a non-simultaneous but resolution-matched spectral index, finding $\alpha_{(15-129)\,\mathrm{GHz}}=-0.03\pm0.23$ between 15-129~GHz.
Such a flat spectrum agrees with the 230~GHz flux lower limit constrained by the EHT observations.
Although the 22~GHz KVN flux appears slightly high compared to the 15~GHz flux, 
we note that this is most likely due to the blending effect we mentioned in Sect. \ref{subsec:result}.
The resolution-matched core flux between 5-22~GHz indeed shows a substantially flat spectrum within $\sim1$~mas from the intensity peak 
(see Fig. 4 of \citealt{hada12}).
Therefore, we conclude that the M\,87 core on a submilliarcsecond scale has 
a substantially flat radio spectrum between 15-129\,GHz and possibly up to 230\,GHz.


\section{Discussion}\label{sec:discussions}

\subsection{Comparison with theoretical predictions}

It has been routinely considered that featureless flat spectra of extragalactic jet sources
are made by inhomogeneous plasma consisting of multiple components with
different turn-over frequencies (i.e., ``cosmic conspiracy''; \citealt{cotton80}).
Our KVN data do not allow for a reliable spectral decomposition 
because they lack spatial resolution.
However, the flat spectrum up to at least 129~GHz suggests
the coexistence of qualitatively different plasma with low ($<$100\,GHz) and high synchrotron peak frequencies ($>$100\,GHz).
We can interpret the latter based on theoretical studies of 
energetics of relativistic plasma near the central engine of M\,87.
This connection is provided by the fact that synchrotron radiation from the core region of M\,87 at short millimeter wavelengths 
originates from the vicinity of its central BH \citep{hada11}.
In this regard, a peak frequency higher than 100~GHz supports the broadly investigated idea about a jet dominated by magnetic energy that is launched at the base of the M\,87 jet \citep{broderick09,kino15,punsly17}. 

The underlying flat spectrum in the intermediate frequency range ($\sim15-86$~GHz) is also interesting.
This cannot be easily explained by the expanding jet 
because the jet has a turn-over frequency below 24~GHz already at a core distance of $\gtrsim0.2$~mas  \citep{hada16}.
Interestingly, \cite{kino15} showed that 
a single, homogeneous, ultra-compact ($\sim40~\mu$as), and fully synchrotron self-absorbed (SSA) plasma at 230\,GHz 
would cause problems for several reasons.
The authors specifically demonstrated that the compact size and the high turn-over frequency imply a magnetic field strength of $B\approx300$\,G.
This value leads to an unrealistically great electromagnetic jet power and extremely short synchrotron cooling timescale,
which contradicts observations (e.g., \citealt{doeleman12}).
Based on this, the authors suggested that the jet base region consists of at least 
two zones with different levels of magnetization,
which could produce another synchrotron peak in the intermediate-frequency regime.

\subsection{Constraining the turn-over frequency of the magnetic energy dominated plasma}

Because we lack direct flux measurements between 129 and 230\,GHz,
it is unclear whether the highest turn-over frequency can be even higher than 230\,GHz or
if the entire jet base becomes optically thin at 230\,GHz.
In the latter case, \cite{kino15} suggested the spectrum turn-over to be at 160\,GHz (see Fig. 5 therein)
and a very steep spectrum ($\alpha\leq-2.5$) between 160 and 230\,GHz.
In order to examine the latter case, we took the 129\,GHz core flux of $\sim0.9$\,Jy from the KVN data
as an upper limit for the 129\,GHz flux on the EHT-scale region ($\lesssim40\mu$as).
We also assumed a flat spectrum between 129 and 160~GHz, but a steep spectrum between 160 and 230~GHz.
This gives the 230~GHz flux upper limit at 230~GHz of 0.9\,Jy$\times(230/160)^{-2.5}\sim0.4$\,Jy.
Even a much more moderate steep spectral index $\alpha=-0.7$ predicts 0.9\,Jy$\times(230/160)^{-0.7}\sim0.7$~Jy at 230~GHz.
These fluxes are not sufficient to explain the 230\,GHz flux measured by the EHT observations.
Therefore, we favor the former scenario.

\section{Summary} \label{sec:summary}

In this letter, we have presented a study of the synchrotron spectrum of the nuclear region in M\,87
by KVN VLBI monitoring observations that have been regularly performed over four years 
at 22, 43, 86, and 129\,GHz.
The KVN observations resolved the core at a projected linear size of $\lesssim0.8$~mas$\sim110R_{s}$.
The KVN data constrained the core spectral index $\alpha\gtrsim-0.37$ between 22-129~GHz.
Other resolution-matched multi-frequency observations
suggest the flat spectrum over a wider range of frequency (15-230~GHz).
The flat spectrum extending up to short millimeter wavelengths
implies a strong magnetization in the jet base, consistent
with the theoretical predictions (e.g., \citealt{broderick09,kino15}).
Therefore, we suggest a magnetically dominated jet-launching scenario for M\,87.

\begin{acknowledgements}
We thank the anonymous referee for the careful reading and valuable comments that helped us to improve the paper.
J.-Y. Kim is supported for this research by the International Max Planck Research School (IMPRS) for 
Astronomy and Astrophysics at the University of Bonn and Cologne.
J.-Y. Kim would like to thank Eduardo Ros, Manel Perucho, and Christian Fromm for fruitful discussions.
S.-S. Lee was supported by the National Research Foundation of Korea (NRF) grant funded by the Korea government (MSIP) 
(No. NRF-2016R1C1B2006697).
G.-Y. Zhao is supported by the Korea Research Fellowship Program through the National Research Foundation of Korea (NRF) 
funded by the Ministry of Science and ICT (NRF-2015H1D3A1066561).
We are grateful to all staff members in KVN who helped to operate the array and to correlate the data. 
The KVN is a facility operated by KASI (Korea Astronomy and Space Science Institute). 
The KVN operations are supported by KREONET (Korea Research Environment Open NETwork) ,
which is managed and operated by KISTI (Korea Institute of Science and Technology Information).
This research has made use of data from the MOJAVE database that is maintained by the MOJAVE team \citep{lister09}.
\end{acknowledgements}

%
%

\bibliographystyle{aa}
\bibliography{m87kvn}


\begin{appendix}

\section{
Simulation of KVN visibilities and estimation of 
a systematic trend in the spectral index
}

The analysis of multi-frequency interferometric images is sensitive to the finite sampling of the source information in the visibility domain (e.g., \citealt{fromm13}).
This issue can be particularly important for the KVN data because of the limited $(u,v)$ coverage of the array 
(see Fig. 2 of \citealt{lee16}).
Therefore, we evaluated the effect of the frequency-dependent $(u,v)$ coverage on the accuracy of the spectral index measurement.

The main goal was to feed the simulator with exactly the same source model with a flat spectrum for different $(u,v)$ coverages 
(i.e., only different observing frequencies) and test if 
an absolutely flat spectral index can be recovered from the simulated data.

\begin{figure}
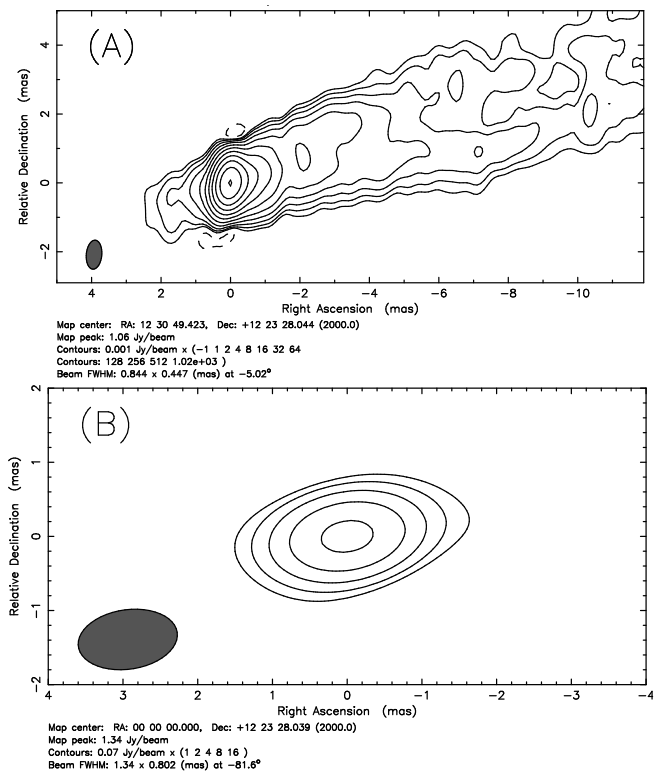

\centering
\includegraphics[width=0.27\textwidth, angle=-90, trim={33pt 0 0 0}, clip]{./figs/m87vlba15g.eps}

\includegraphics[width=0.28\textwidth, angle=-90, trim={33pt 0 0 0}, clip]{./figs/m87kvn86g_sim.eps}
\caption{
{(A)} : Input M\,87 clean model image obtained by the VLBA observations at 15\,GHz
{(B)} : KVN 86~GHz image obtained from a simulated 2hr long KVN snapshot observation
of the model image in panel (A). 
}
\label{fig:appendix1}
\end{figure}

For this purpose, we prepared a ground-truth image of the M87 jet to simulate synthetic KVN observations.
A full-track M87 VLBA 15\,GHz data set was taken from the MOJAVE database (VLBA code BK145A),
and we imaged the source structure using the \textsc{Difmap} task \textsc{clean}.
The maximum baseline length of this observation was 440\,M$\lambda$
and provided an angular resolution of $0.45\times0.84$\,mas.
In panel (A) of Fig. \ref{fig:appendix1} we show the corresponding inner jet structure.
This clean model was imported into the NRAO \textsc{AIPS} software \citep{aips}
to be used as an input for the \textsc{AIPS} task \textsc{UVCON}.
Then, we generated synthetic datasets using exactly the same input clean model for the four different KVN observing frequencies.
The entire simulation was performed by following several specific steps.

\begin{enumerate}
 
 \item A full-track synthetic data set was created by using \textsc{AIPS UVCON}.
 We assumed
 one of the KVN observing frequencies (e.g., 22\,GHz), 
 the KVN station coordinates,
 and a 10\,hr observing run. 
 System temperatures and aperture efficiencies were chosen to be close to typical values in real observations.
 We note that the decoherence was not considered in the simulation.

 \item From session to session, KVN observed M\,87 in snapshot mode at different baseline position angles.
 In order to account for this, we split the simulated full-track data into five subsets,
 each of which had the same 2~hr of on-source time (e.g., LST 6:00-8:00, 8:00-10:00, 10:00-12:00, 12:00-14:00, and 14:00-16:00).

 \item Each sub-dataset was loaded into \textsc{Difmap,} and we 
 estimated the core flux by fitting a circular Gaussian model to the visibilities, as described in Sect. \ref{sec:method}
 (see panel (B) of Fig. \ref{fig:appendix1}).
 
 \item The core flux estimate was repeated for all the five sub-datasets.
 Then, we took the average and standard deviation of the five different flux values
 to determine a characteristic flux value.
 
 \item Steps 1 to 4 were repeated for the synthetic data sets at observing frequencies of 22, 43, 86, and 129~GHz.
 
\end{enumerate}

\begin{figure}
\centering
\includegraphics[width=0.45\textwidth]{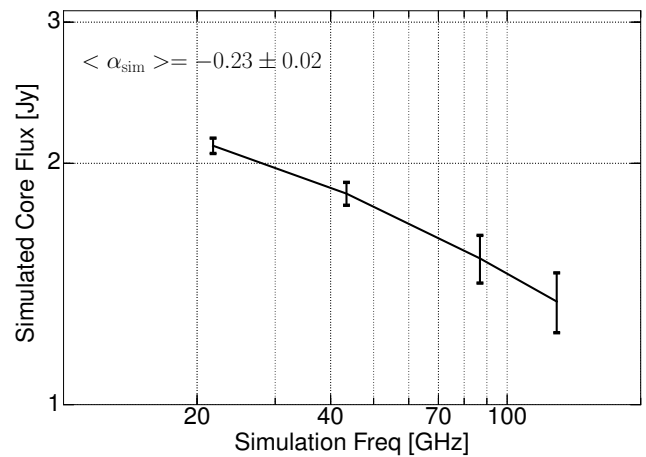}
\caption{
Artificial M\,87 core flux versus the simulation frequency obtained by the simulated observations.
}
\label{fig:appendix2}
\end{figure}

In Fig. \ref{fig:appendix2} we show the artificial spectrum obtained by the simulation.
We were not able to recover a completely flat spectral index and 
rather obtained $\alpha_{\rm sim}$ of $-0.23\pm0.02$.
Most likely, this is due to different level of the array sensitivity to extended emission at limited uv spacings.
Therefore, the limited (u,v) coverage very likely causes similar systematic effects in the analysis of the real multi-frequency M\,87 data as well.

After these considerations, we conclude that it is unlikely that the actual source spectrum is even slightly steep or inverted in our observing band.

\end{appendix}

\end{document}